\documentclass[aip,apl,preprint,groupedaddress]{revtex4-1}
\pdfoutput=1
\usepackage{array, makecell}
\usepackage{dcolumn}   
\usepackage{bm}        
\usepackage{amssymb}   
\usepackage{amsfonts}
\usepackage{verbatim}
\usepackage[pdftex]{graphicx}  
\usepackage{amsmath}
\usepackage{amsfonts}
\usepackage{ulem}
\usepackage[pdftex]{color}

\begin{document}
\normalem

\title{High-resolution Compton spectroscopy using X-ray microcalorimeters}

\author{U. Patel$^{1}$}
\email[Electronic address: ]{upatel@anl.gov}
\author{T. Guruswamy$^{1}$}
\author{A. J. Krzysko$^{1}$}
\author{H. Charalambous$^{1}$}
\author{L. Gades$^{1}$}
\author{K. Wiaderek$^{1}$}
\author{O. Quaranta$^{1}$}
\author{Y. Ren$^{1}$}
\altaffiliation[Present address: ]{Department of Physics, City University of Hong Kong, Kowloon, Hong Kong, China.}
\author{A. Yakovenko$^{1}$}
\author{A. Miceli$^{1}$}

\affiliation{$^{1}$X-ray Science Division, Argonne National Laboratory, Lemont, IL 60439, USA}

\date{\today}

\begin{abstract}
X-ray Compton spectroscopy is one of the few direct probes of the electron momentum distribution of bulk materials in ambient and $\textit{operando}$ environments. We report high-resolution inelastic X-ray scattering experiments with high momentum and energy transfer performed at a storage-ring-based high-energy X-ray light source facility using an X-ray microcalorimeter detector. Compton profiles were measured for lithium and cobalt oxide powders relevant to lithium-ion battery research. Spectroscopic analysis of the measured Compton profiles shows high-sensitivity to the low-$Z$ elements and oxidation states. The lineshape analysis of the measured Compton profiles in comparison with computed Hartree-Fock profiles is limited by the resolution of the energy-resolving semiconductor detector. We have characterized an X-ray transition-edge sensor microcalorimeter detector for high-resolution Compton scattering experiments using a bending magnet source at the Advanced Photon Source (APS) with a double crystal monochromator providing monochromatic photon energies near 27.5\,keV. The momentum resolution below 0.16\,atomic units was measured yielding an improvement of more than a factor of 7 over a state-of-the-art silicon drift detector for the same scattering geometry. Furthermore, the lineshapes of narrow valence and broad core electron profiles of sealed lithium metal were clearly resolved using an X-ray microcalorimeter detector compared to smeared and broadened lineshapes observed when using a silicon drift detector. High-resolution Compton scattering using the energy-resolving detector shown here presents new opportunities for spatial imaging of electron momentum distributions for a wide class of materials with applications ranging from electrochemistry to condensed matter physics.
\end{abstract}

\maketitle

\begin{figure}[t]
\includegraphics[width=.48\textwidth]{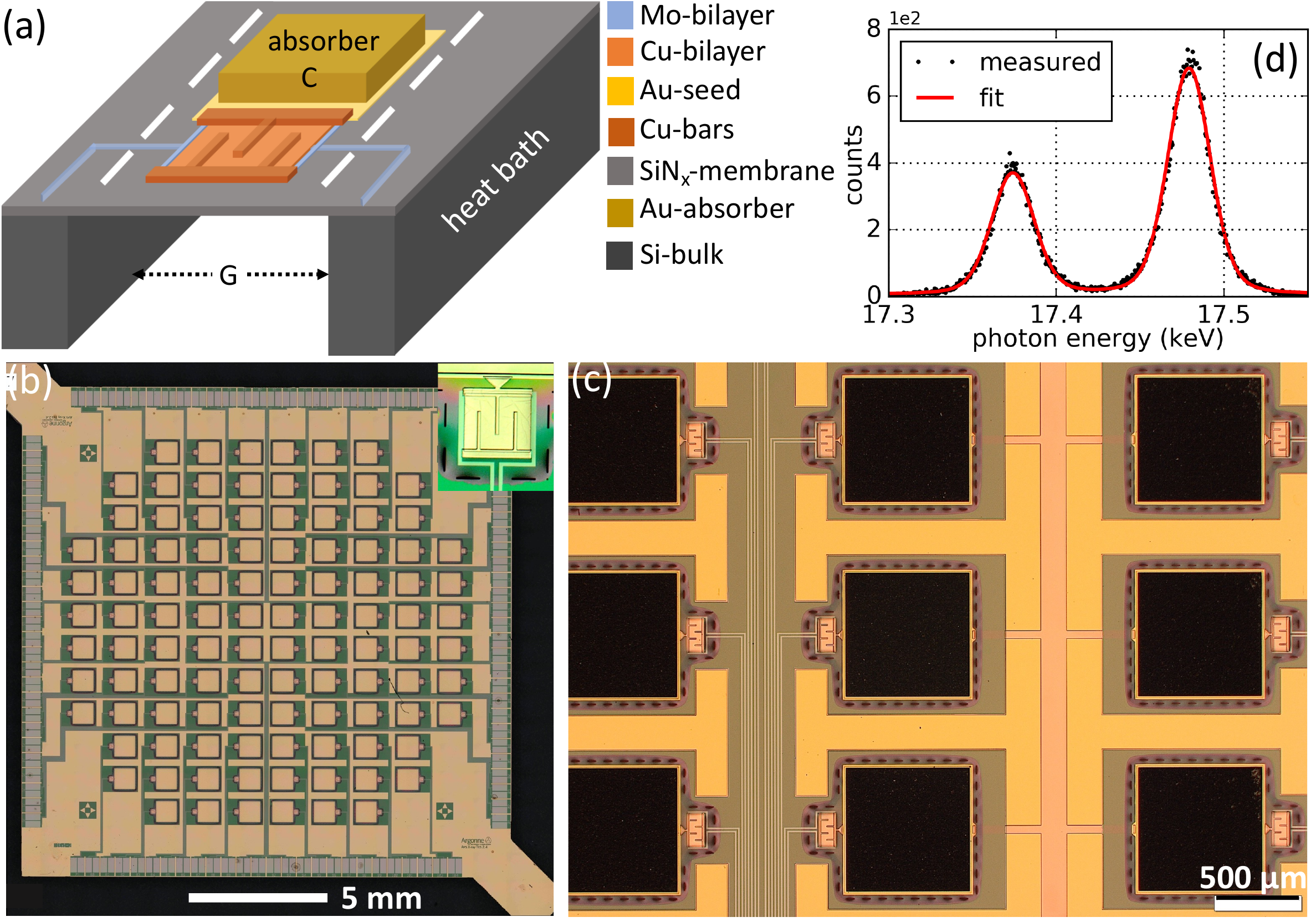}
\vspace{-10.0pt} \caption{(a) 3D view of the completed device showing color coded layer stackup for a TES pixel (not-to-scale). Optical micrographs of the fabricated sensor chips showing (b) 100-pixel array with $\sim\,1\,\mu$m thick sputtered Au absorber and (c) scalable high-energy TES pixels with electroplated metal stack of $\sim\ 2\,\mu$m thick Au and $\sim\ 22\,\mu$m thick Bi absorbers incorporated in the TES fabrication process. (d) Prototype device characterization of high-energy TES design (absorber: 2\,$\mu$m thick Au) using a lab X-ray tube source showing measured Mo K$\alpha$ line and corresponding fit for estimating detector energy resolution.}
\vspace{-12.0pt} 
\label{fig:figure1}\end{figure}

X-ray Compton scattering is a powerful technique for directly probing ground state electron momentum distributions for a wide class of materials under a variety of experimental conditions. The Compton technique offers several immediate advantages over other momentum (\textbf{k})-resolved spectroscopies: high X-ray energy allows for deep penetration into encased samples and the probe is bulk-sensitive but insensitive to crystal defects and surface effects. \cite{Cooper04} These characteristics allow for studies of many complex electronic structure material systems. Compton scattering spectroscopy has been a valuable tool for studying charge transfer \cite{Chabaud04} and extracting redox orbitals \cite{Suzuki15} in Li battery cathode materials as well as orbital character of dopants \cite{Sakurai11} and bulk Fermi surface \cite{Sawai12,Dugdale06,Hiraoka20,Sakurai95} in a broad class of condensed matter systems. The inelastic scattering experiments in the Compton limit are typically performed with a combination of a strong synchrotron X-ray source and energy or crystal dispersive analysis of the scattered radiation. High-resolution in momentum space along with high-energy and high-flux is necessary for full realization of the capabilities of the Compton spectroscopic technique. In the Compton regime, within the limits of the impulse approximation, the measured energy spectrum is directly proportional to the Compton profile: $J(p_z)\,=\,\iint \rho (\textbf{p})dp_xdp_y\,$, where $\rho(\textbf{p})$ is the electron momentum density with scattering vector aligned in the direction of the $p_z$-axis. Thus, by energy-analyzing the Compton scattered radiation using high-resolution detectors, it is possible to directly access information about the electronic ground state of the scattering system. Commercial Ge detectors are typically used for this application due to their high quantum efficiency at high X-ray energy ($>$\,30\,keV). However, use of such detectors is often limited due to their low energy (momentum) resolution, typically $\sim$\ 500\,eV\ (0.55\,atomic units (a.u.))\ at 100\,keV. The energy resolution of semiconductor (Ge, Si) detectors are typically limited by charge generation statistics within the detector which is given by $\Delta E \sim 2.35 \sqrt{\epsilon F E}$, where $\epsilon$ is the pair creation energy, $F$ is the Fano factor and $E$ is the photon energy. \cite{Knoll10} 

For calorimetric X-ray sensor designs, the thermodynamic limit: $\Delta E \sim 2.35 \sqrt{k_BT^2C}$ is much lower compared to the theoretical limit for Ge or Si. Although early progress was reported for Compton scattering measurements using a low temperature detector based on a silicon thermistor,\cite{Stahle92} good energy resolution and scalability needed for high counting efficiency remains a challenge. X-ray transition-edge sensor (TES) microcalorimeter detectors offer energy resolution of $\sim$\ 50\,eV at 100\,keV and 250-pixel array count rates up to 2.5\,kcps, providing an order of magnitude improvement. \cite{Ullom15} Low-$Z$ elements such as Li are typically hard to detect with X-ray fluorescence (XRF) measurements due to their low energy levels giving a weak fluorescence signal. On the other hand, Compton signal is relatively strong from low-$Z$ elements, enabling non-destructive light element imaging in battery cells. Moreover, scanning Compton X-ray microscopy, with image resolution approaching tens of nanometers, using hard X-ray detection technology with near 4$\pi$ solid angle coverage has recently been proposed and implemented for cellular imaging of biological and radiosensitive samples. \cite{Perez21} There have also been successful efforts reported from SPring-8 for quantification of Li for $\textit{in-situ}$ and $\textit{operando}$ lithium batteries using Ge detectors; the high energy photons, up to 100\,keV, allow deep penetration. \cite{Suzuki16,Suzuki19} For imaging purposes, Compton profiles with relatively low resolution and statistics are typically collected. However, Compton lineshape analysis is difficult to perform due to significant smearing and broadening. Conversely, the high-resolution Compton spectrum collected using Cauchois type crystal spectrometer with resolution $\sim$\ 0.14\,a.u.\ was successfully used to study electronic states in cathode materials of Li batteries during lithiation and delithiation. \cite{Hafiz21,Hafiz17} 

\begin{figure}
\includegraphics[width=.48\textwidth]{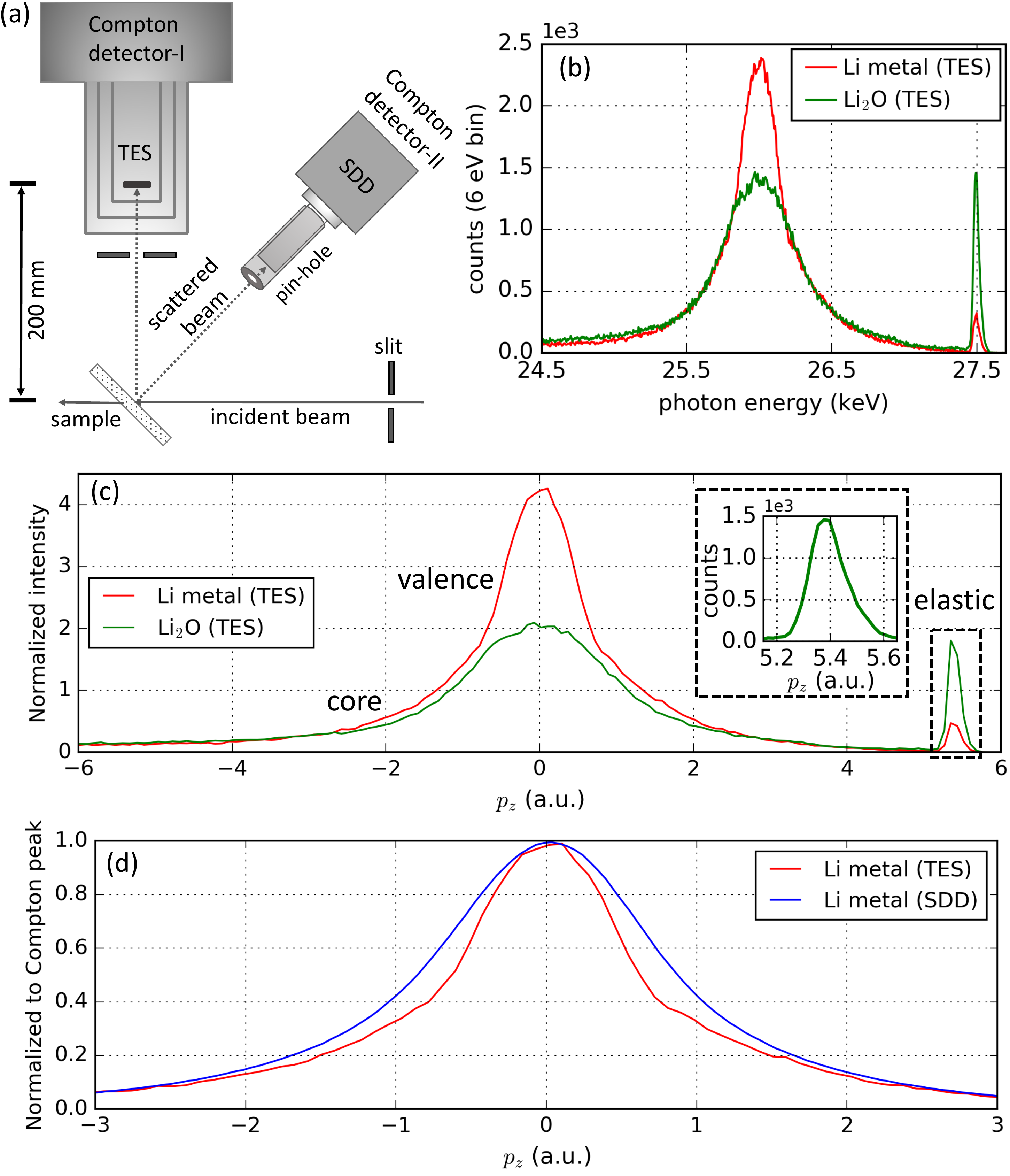}
\vspace*{-10.0pt} \caption{Characterization of X-ray TES microcalorimeters for lithium detection and high-resolution lineshape analysis. (a) Schematic of an experimental setup at the 1-BM-C beamline showing the TES instrument fixed with respect to the beam and Hitachi Vortex silicon drift detector (SDD) with a pin-hole collimator. TES array is mounted at the cold stage typically held at 50\,mK. The scattered photons, in a narrow solid angle defined by a slit, pass through several X-ray windows before getting detected by the sensor. Each pixel measures the Compton signal. (b) Raw counts of scattered photon energy resolved using TES for Li metal and Li$_{2}$O. (c) Normalized-to-wings ($\lvert p_z \rvert$\,=\,2 to 6\,a.u.\,) Compton spectrum of Li metal and Li$_{2}$O measured with TES showing high-sensitivity of detecting Li concentration. Inset: narrow elastic line monitors the instrument resolution. (d) Normalized-to-peak Compton spectrum of Li metal measured with TES showing clear non-gaussian lineshape corresponding to superposition of valence and core electron profiles, while SDD poor detector resolution caused broadening and smearing of lineshape.}
\vspace*{-12.0pt}
\label{fig:figure2}\end{figure}

In this letter, we demonstrate high-resolution Compton spectroscopy using hard X-ray TESs fabricated and tested at Argonne National Laboratory (ANL) and analyze sensitivity of Compton profile for detecting low-$Z$ elements, particularly lithium. Furthermore, we evaluate these sensors for lineshape analysis, which is necessary for studying orbitals related to fundamental redox processes in lithium batteries. A 3D physical layout of a TES microcalorimeter for a completed device is shown in Fig.~\ref{fig:figure1}(a). The sensor (Fig.~\ref{fig:figure1}(b)) is composed of a square (125\,$\mu$m)$^{2}$ proximity-coupled superconducting Mo/Cu thin-film bilayer with a ``sidecar" normal-metal Au absorber with area (750\,$\mu$m)$^{2}$. The bilayers were deposited using optimal sputter deposition parameters so that the thin-film internal stress in Mo is minimized from a few-GPa tensile stress to slightly compressive stress ($<$\,100\,MPa). The resultant bilayer films have a good superconducting state with a sharp transition temperature of 90\,mK, and a normal state sheet resistance of 8\,m$\Omega$/$\square$. The X-ray absorber and thermometer were deposited on a 0.5-$\mu$m-thick suspended silicon nitride (SiN$_x$) membrane which was released by a bulk silicon micromachining method. The heat bath was cooled below 50 mK using an adiabatic demagnetization refrigerator. We have tested sensors with 1\,$\mu$m thick sputtered Au absorber, a subset of pixels from a 100-pixel array, with measured energy resolution below 20\,eV for X-ray energies up to 20\ keV. The measured thermal conductance, G, and heat capacity, C, were $\sim$\ 550\,pW/K and $\sim$\ 4\,pJ/K respectively, close to the target values. The measurements were consistent with those reported for a different 24-pixel array design.\cite{Guruswamy21} In Fig.~\ref{fig:figure1}(c), a high-energy TES design with electroplated Au and Bi absorber fabricated at ANL, is shown. The heat capacity expected for this design is 8.7\,pJ/K, which increases the saturation energy ($E_{sat} \propto C/\alpha$, where $\alpha \propto$ 1/transition-width) into the deep hard X-ray regime.

\begin{table}
\vspace*{-6.0pt}\caption{\label{tab:table4}Estimated experimental momentum resolution of detectors used: TES ($\phi$ = 90$^\circ$) and SDD ($\phi$ = 140$^\circ$).}
\begin{ruledtabular}
\begin{tabular}{ccccc}
Detector&
\makecell{$dp_z/d\phi$\\(a.u./rad)}&
\makecell{$dp_z/dE_2$\\(a.u./keV)}&
\makecell{$dp_z/dE_1$\\(a.u./keV)}&
\makecell{$\Delta p_z$\\(a.u.)}\\
\hline
TES&5.1& 3.81& 3.43 & 0.18 \\
SDD&2.41&3.03& 2.53 & 0.92\\
\end{tabular}
\end{ruledtabular}
\vspace*{-12.0pt}
\label{table:table1}
\end{table}

In Fig.~\ref{fig:figure2}(a), the Compton scattering geometry at the 1-BM beamline for high-resolution inelastic scattering experiments using TES instrument is presented; a 100-pixel sensor chip was packaged and installed at the cold stage. The sample under study was set at 45$^\circ$ with respect to the incident beam at 27.5\,keV. The size of the X-ray beam interacting with the sample was reduced to 5\ $\times$\ 3\,mm$^{2}$ (H$\,\times\,$V) by closing incident beam slits. Scattered X-ray photons were counted and energy-resolved by two Compton detectors at two different scattering angles: TES detector at 90$^\circ$ (perpendicular geometry) and SDD at 140$^\circ$ (backscatter geometry). The two geometries allow direct comparison of the data from TES and SDD for scattering geometries favorable for imaging and lineshape analysis, respectively. Typically, the lineshape and width of the Compton profile are directly related to the electron orbitals, whether the electrons are in the valence or the core shells. However, further contributions to the broadening of the profile can be associated with instrumental and geometric conditions: $\Delta\,p_z\,=\,\sqrt{\left(\frac{\partial p_z}{\partial E_1}\Delta E_1\right)^2 + \left(\frac{\partial p_z}{\partial E_2}\Delta E_2\right)^2 + \left(\frac{\partial p_z}{\partial \phi}\Delta \phi\right)^2}$, where $E_{1}$ and $E_{2}$ are the incident and scattered photon energies, respectively, and $\phi$ is a scattering angle. Thus, it is important to minimize different contributions adversely broadening the Compton spectrum. The source broadening $\Delta E_1$ is the energy spread of the incident beam which is roughly proportional to the Darwin width of the Si (111) monochromator crystal. For the 1-BM experiments using a bending magnet source, the energy spread of the incident X-ray beam was $\Delta E_1/E_1\,$$\sim$\,$1.5\times10^{-4}$ with a typical flux $>$\ 10$^{10}$ photons/sec for energies below 30\,keV. Aluminum slits and brass pinhole collimators were used to limit the broadening due to the spread in scattering angle ($\Delta\phi$) while maximizing the number of photons collected for the TES and SDD, respectively. In Table~\ref{table:table1}, estimates for the total experimental momentum resolution for the scattering measurements are listed. To prevent oxidation induced broadening of the profile, air sensitive samples, particularly lithium, were carefully sealed in an Ar-filled glovebox with $<$ 0.1\,ppm oxygen and $<$ 0.5\,ppm water in a polyethylene pouch. Containers were chosen to ensure low incident and scattered beam attenuation and minimal background contribution. Lithium metal enables evaluation of the high-energy TES microcalorimeters due to its inherently narrow Compton profile. 

\begin{figure}[t]
\includegraphics[width=.48\textwidth]{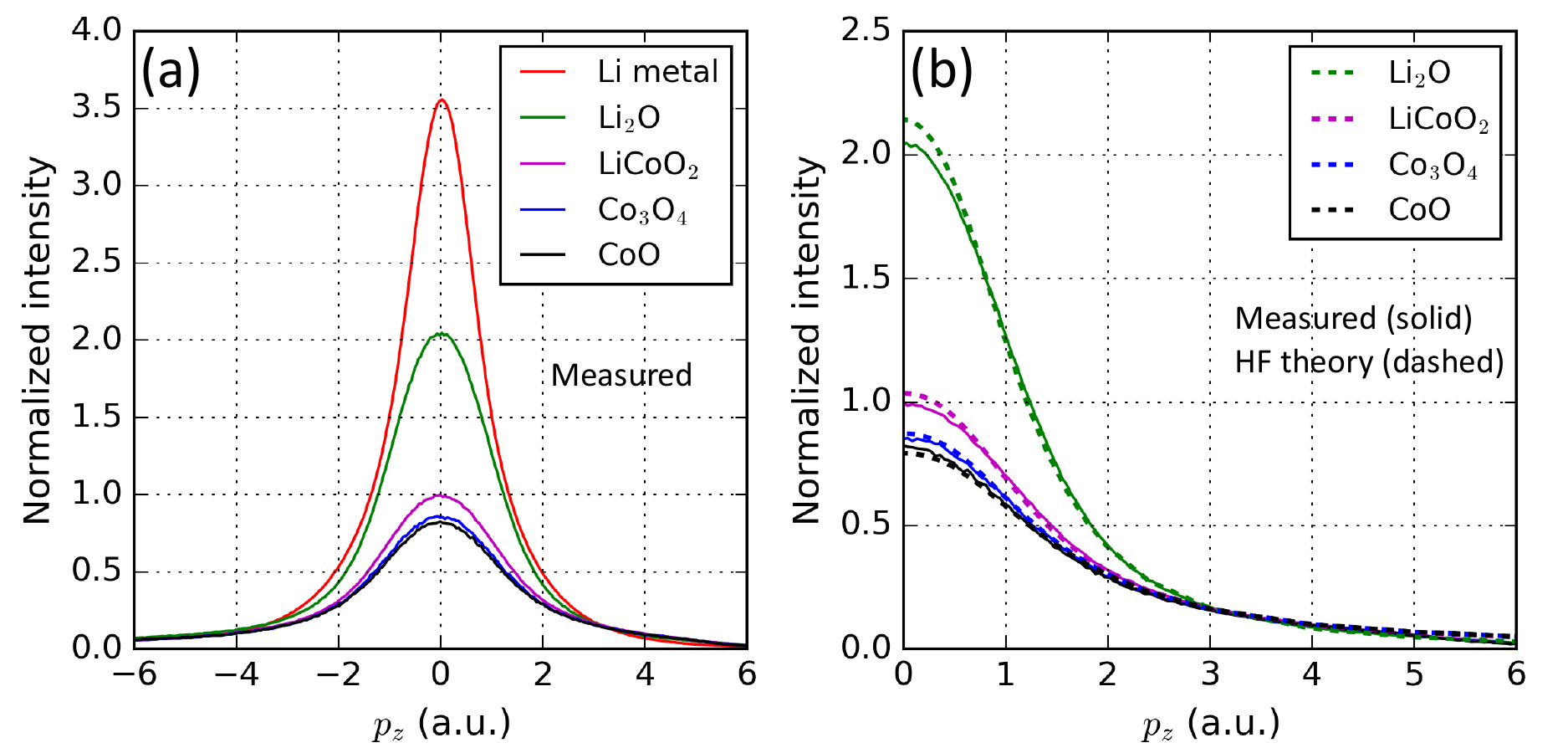}
\vspace*{-10.0pt} \caption{SDD-measured and computed Compton profiles of the battery materials: (a) The normalized Compton profiles for Li, Li$_{2}$O, LiCoO$_{2}$, Co$_{3}$O$_{4}$ and CoO showing high-sensitivity to the low-$Z$ elements for which the electron momentum distribution is narrow and focused around $p_z$\ =\ 0. (b) Comparison of the measured (solid) Compton profiles for the data shown in (a) and computed (dashed) Compton profiles using HF theory. \cite{Biggs75} Oxidation states of Li and Co were used for the computed profiles with the charge transfer to O (2$p$) orbitals.}
\vspace*{-12.0pt} 
\label{fig:figure3}\end{figure} 

\begin{figure}[t]
\includegraphics[width=.48\textwidth]{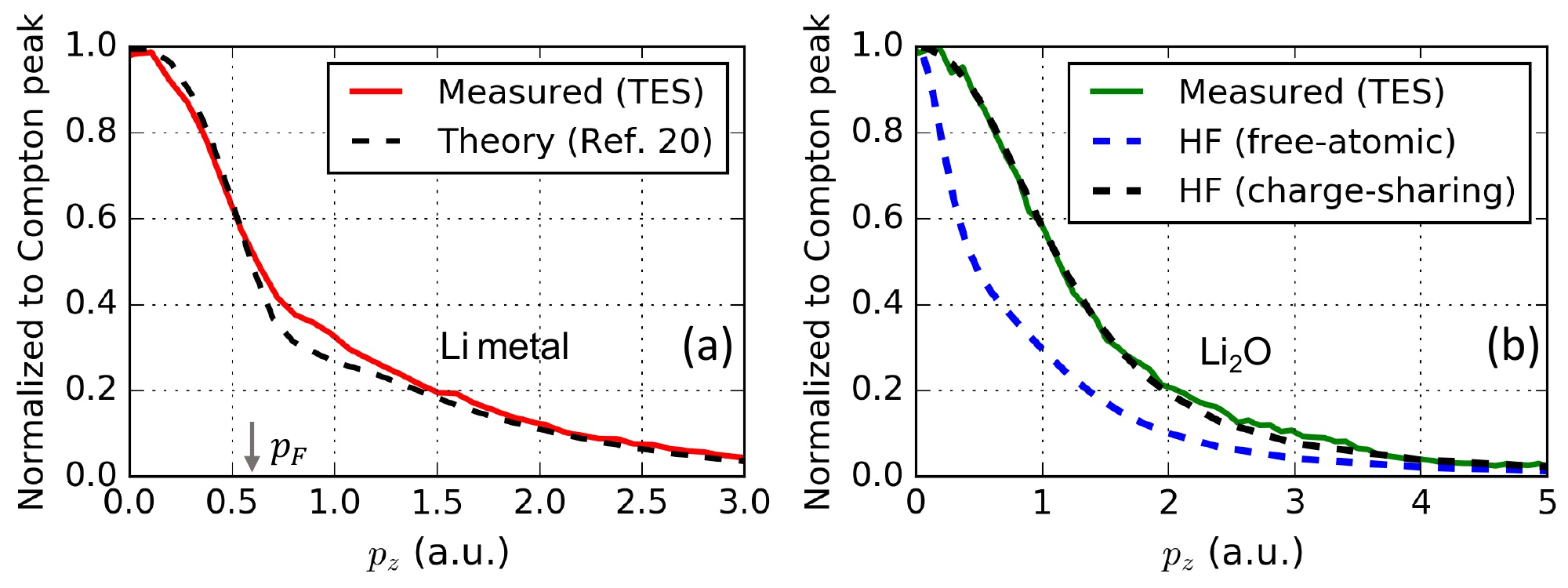}
\vspace*{-10.0pt} \caption{(a) Measured and theoretical \cite{Dovesi83} total Compton profiles for Li metal. The theoretical profile was convoluted with the experimental resolution.\ $p_F$ indicates the Fermi momentum for Li metal. (b) Comparison of the Compton profile measured with TES and the computed weighted sum of atomic profiles of lithium and oxygen using HF theory \cite{Biggs75} for Li$_2$O with and without charge sharing in outer orbitals.}
\vspace*{-12.0pt} 
\label{fig:figure4}\end{figure}

Compton profiles were measured for lithium metal, lithium oxide, lithium cobalt(III) oxide, cobalt(II, III) oxide and cobalt(II) oxide powders (Li, Li$_{2}$O, LiCoO$_{2}$, Co$_{3}$O$_{4}$, CoO), relevant to lithium-ion battery research. Figure~\ref{fig:figure2}(b) shows the raw energy spectra from Li metal and Li$_{2}$O measured with TES. The measured energy spectra were then converted to the electron momentum scale using the relation\cite{Cooper04}: $\frac{p_z}{mc} = \frac{E_2-E_1+(E_2E_1/mc^2)(1-cos\phi)}{\sqrt{E_1^2 + E_2^2 - 2E_1E_2cos\phi)}}$, where m is the electron mass and c is the speed of light. The area under the Compton profile subjects to the normalization rule:$\int^{+\infty}_{-\infty} J(p_z) dp_z\,=\,Z\,$. Figures~\ref{fig:figure2}(c) and \ref{fig:figure3}(a) show measured Compton profiles of the battery materials using TES and SDD, respectively. As outlined by the S-parameter approach, \cite{Suzuki16} all of the measured profiles were then normalized to the wings ($\lvert p_z \rvert$\,=\,2 to 6\,a.u.\,) to identify the sharpness of the distributions. The amplitudes of the resulting Compton profiles clearly show high detection sensitivity, particularly for low-$Z$ elements such as lithium for which the electron momentum distribution is narrowly concentrated around $p_z$\ =\ 0 (primarily, $\lvert p_z \rvert$\,=\,0 to 2\,a.u.\,for Li). Compton profiles with broad distributions, with small change in amplitudes, were also observed for the two stable oxidation states of cobalt(III) and cobalt(II, III) oxide. In Fig.~\ref{fig:figure3}(b), the comparison of SDD-measured and theoretical Compton profiles show good agreement with Hartree-Fock (HF) theory. \cite{Biggs75} However, simultaneous measurements with the TES array show that the widths, amplitudes and lineshapes of the Compton profiles collected with the SDD are limited by the detector resolution, hindering the ability to collect profiles that accurately display the electronic structures intrinsic to the materials and also reducing the element-resolving Compton contrast for imaging. As shown in Fig.~\ref{fig:figure2}(c), the Compton profile lineshape from lithium metal was clearly resolved by the TES detector, showing a clear slope change between narrow valence and broad core electron profiles. As an approximation, using the free-electron gas model for Li metal: $J_{2s}(p_z) = \{3(p_F^2 - p_z^2)/4p_F^3$ for $p_z \leq p_F$; 0 for $p_z > p_F$\}, where $p_F$ is the Fermi momentum. The measured lineshape shows expected inverted parabola with two inflection points on both sides of the Compton profile. This indicates a cross-over where the dominant contribution to the Compton profile shifts from valence electrons to core electrons. Valence electron profile with inverted-parabola shape was clearly observed using the TES microcalorimeter due to its high-resolution. High resolution also enabled separating out the elastic peak from the Compton peak, which would otherwise be difficult when using SDD for a perpendicular geometry with a relatively low incident beam energy of 27.5\,keV. For the elastic peak, the measured FWHM momentum (energy) resolution with SDD was 0.91\,a.u.\ ($\sim$\ 300\,eV) for backscattering geometry which corresponds to 1.15\,a.u.\ for the perpendicular geometry. For the SDD, in addition to charge generation statistics ($\epsilon$~$\sim$\ 3.63\,eV, $F$\ $\sim$\ 0.135, $\Delta E$\ $\sim$\ 273\,eV), \cite{Perotti99} other components including charge carrier collection and electrical noise affect the overall resolution via quadrature sum. For the TESs, the measured resolution was below 0.16\,a.u., which is comparable to using a crystal analyzer for the same scattering geometry and a factor of 7 better than when using SDD. An additional factor of 2 improvement is within reach of the sensor and readout capability and by further reduction in $\Delta \phi$. In Fig.~\ref{fig:figure2}(d), smearing and gaussian broadening of the lineshape is visually demonstrated for the SDD, but a clearly resolved lineshape is evident when using TESs. Due to detector resolution convolution, the FWHM of the measured lithium metal Compton profile with the SDD was $\sim$\ 1.8\,a.u.\ compared to 1.2\,a.u., close to the intrinsic width, measured with TES which is comparable to the directional profiles collected using crystal analyzers. \cite{Sakurai95} In Fig.~\ref{fig:figure4}(a), total Compton profile measured with TES is presented for Li, which is in close agreement to the resolution-broadened theoretical HF profile with correlation corrections. \cite{Dovesi83} The discrepancy could be attributed to the multiple scattering events and residual oxidation in Li. In Fig.~\ref{fig:figure4}(b), a comparison is presented for normalized Compton profiles for Li$_2$O measured with TES, computed HF neutral atom and HF total orbitals with charge sharing between lithium (2$s$) and oxygen (2$p$) orbitals based on its oxidation state. Excellent agreement was found  between the measured profile using TES and the HF theory incorporating charge sharing. Further developments of the sensor design with high $E_{sat}$, initial characterization shows measured energy resolution below 30\,eV for the Mo K$\alpha$ (17.5\,keV) line (Fig.~\ref{fig:figure1}(d)); the expected benchmark performance has recently been reported for high-energy designs with C up to 7\,pJ/K obtained with evaporated Au. \cite{Yan21} Thus, a high-energy TES instrument at a high-energy ($>$\ 100\,keV) beamline serves as a microscopic probe to measure redox orbitals induced changes in $\textit{operando}$ batteries and potentially as an imaging tool after scaling up array size of the TES sensor and cold readout circuits. Hence, it would open the door to new experiments not only for Compton imaging of light elements but also for high-resolution X-ray spectroscopies for the entire energy range of 20-100\,keV, including monitoring physical phenomena such as metal to insulator transition via spherically averaged Compton signal, \cite{Ruotsalainen18,IIkka19} and Fermi surface structures in bulk materials. \cite{Sawai12,Hafiz17} After the APS upgrade, experiments will further benefit from highly focused beams with flux and brilliance necessary to increase counts in a narrow solid angle needed for high-resolution measurements.

In summary, we have demonstrated use of the X-ray TES as a Compton detector for high-resolution inelastic scattering experiments. The X-ray sensors were evaluated for energies up to 27.5\,keV for a perpendicular scattering geometry favorable for imaging and a backscattering geometry favorable for high-resolution lineshape analysis. The X-ray TES improves the momentum resolution by more than a factor of 7, yielding an improvement in measuring Compton profiles even when using low energies and unfavorable scattering geometries. The high-resolution Compton spectroscopy using TESs showed high-sensitivity to the low-$Z$ elements in battery materials, and the measured lineshapes were in good agreement with the computed HF profiles. We further demonstrated smearing of lineshape for the measured Compton profile of lithium metal when using SDD but clear narrow lineshapes of valence and core electron profiles observed when using TES. The momentum resolution is anticipated to further improve for high-energy sensors at high-energy scattering beamlines. The Compton statistics are also expected to improve as our TES pixels and readout packaging grow in array size, enabling highly efficient scattering experiments.

\begin{acknowledgements}
This work was supported by the Laboratory Directed Research and Development program at Argonne National Laboratory. This research used resources of the Advanced Photon Source and Center for Nanoscale Materials, U.S. Department of Energy, Office of Science User Facilities operated for the DOE Office of Science by the Argonne National Laboratory under Contract No.\ DE-AC02-06CH11357. This work made use of the Pritzker Nanofabrication Facility of the Institute for Molecular Engineering at the University of Chicago, which receives support from Soft and Hybrid Nanotechnology Experimental Resource (NSF ECCS-2025633), a node of the National Science Foundation's National Nanotechnology Coordinated Infrastructure. The authors would like to thank M. Wojcik for assistance with the beamline setup, R. Divan for advice on Bi electroplating, T. Cecil and P. Duda for discussions on DRIE method, members of the Quantum Sensors Group, NIST (Boulder, CO USA), for discussions on TES fabrication and cold readout circuits, and U. Ruett for the discussions on Compton scattering experiments.
\end{acknowledgements}

\section*{Data Availability Statement}
The data that support the findings of this study are available from the corresponding author upon reasonable request.


\end{document}